\documentclass[journal = nalefd, manuscript = letter]{achemso}
\usepackage{amsmath,amssymb,bm}
\usepackage{chemformula} % Formula subscripts using \ch{}
\usepackage{physics}
\usepackage[labelsep=period,labelfont=bf]{caption}
\usepackage[capitalise]{cleveref}
\usepackage{sectsty} % for setting fontsize of section
\sectionfont{\fontsize{14}{17}\selectfont}
\usepackage{titlesec}
\titlespacing*{\section}{0pt}{*5.5}{*1}

\newcommand{\numberthis}{\addtocounter{equation}{1}\tag{\theequation}} % for numbering align*

\title{Topological electronic states in holey graphyne}

\author{Yong-Cheng Jiang}
\affiliation{Research Center for Materials Nanoarchitectonics (MANA), National Institute for Materials Science (NIMS), Tsukuba 305-0044, Japan}
\alsoaffiliation{Graduate School of Science and Technology, University of Tsukuba, Tsukuba 305-8571, Japan}
\author{Toshikaze Kariyado}
\author{Xiao Hu}
\affiliation{Research Center for Materials Nanoarchitectonics (MANA), National Institute for Materials Science (NIMS), Tsukuba 305-0044, Japan}
\alsoaffiliation{Graduate School of Science and Technology, University of Tsukuba, Tsukuba 305-8571, Japan}
\email{HU.Xiao@nims.go.jp}

\begin{document} 

  \begin{abstract}
    We unveil that the holey graphyne (HGY), a two-dimensional carbon allotrope where benzene rings are connected by two \ch{-C+C\bond{sb}} bonds fabricated recently in a bottom-up way, exhibits topological electronic states. Using first-principles calculations and Wannier tight-binding modeling, we discover a higher-order topological invariant associated with $C_2$ symmetry of the material, and show that the resultant corner modes appear in nanoflakes matching to the structure of precursor reported previously, which are ready for direct experimental observations. In addition, we find that a band inversion between emergent $g$-like and $h$-like orbitals gives rise to a nontrivial topology characterized by $\mathbb{Z}_2$ invariant protected by an energy gap as large as 0.52~eV, manifesting helical edge states mimicking those in the prominent quantum spin Hall effect, which can be accessed experimentally after hydrogenation in HGY. We hope these findings trigger interests towards exploring the topological electronic states in HGY and related future electronics applications.
  \end{abstract}

\newpage

  Haldane was the first to notice that a typical topological state, i.e., quantum Hall effect, can be realized without Landau levels induced by an external magnetic field. He developed a tight-binding~(TB) model on honeycomb lattice with complex next-nearest-neighbor hoppings which break time-reversal symmetry~(TRS) but introduce no net external magnetic field, generating quantum \textit{anomalous} Hall effect characterized by a Chern number~\cite{Haldane1988}. Although this toy model was thought to be unrealistic at the time of the original proposal, soon after the discovery of graphene~\cite{CastroNeto2009} it was pointed out by Kane and Mele that if spin-orbit coupling~(SOC) is taken into account, graphene can be regarded as two copies of Haldane model, one for up spin and the other for down spin, manifesting the TRS-preserved quantum \textit{spin} Hall effect~(QSHE) characterized by a $\mathbb{Z}_2$ invariant~\cite{Kane2005,Kane2005b} (see also Ref.~\citenum{Bernevig2006}). However, the predicted QSHE is extremely hard to observe experimentally in graphene because of its very weak SOC. In order to resolve this issue, honeycomb systems with heavy elements have been investigated, aiming at having stronger SOC. Recently large-gap QSH insulators have been realized, in Bismuthene~\cite{Reis2017}, monolayer 1T'-\ch{WTe2}~\cite{Wu2018}, stanene~\cite{Deng2018} and ultrathin \ch{Na3Bi}~\cite{Collins2018} for instance. For monolayer 1T'-\ch{WTe2} with an energy gap $\sim$55~meV~\cite{Tang2017}, the quantized edge conductance was observed in a transport channel of 100~nm up to 100~K~\cite{Wu2018}; while for the other materials, although energy gaps range from 0.3~eV to 0.8~eV, evidence for nontrivial topology is still limited in showing the local density of states~(LDOS) for edge states, partially due to their small sample sizes below 40~nm. Despite all the noticeable progresses, further investigations are needed on materials possessing sizable energy gaps and amenable to fabrication into structures for device applications.

  Besides utilizing strong SOC, deforming honeycomb lattice while preserving $C_{6v}$ crystalline symmetry is also able to open a sizable gap in the Dirac dispersion associated with graphene structure, where the band inversion between bands with $p$- and $d$-like characters generates nontrivial topology~\cite{Wu2015,Wu2016,Kariyado2017}. This idea has been experimentally verified first in photonic crystals~\cite{Yang2018,Li2018,Barik2018,Parappurath2020,Wang2023} due to their comparably easy fabrication, and later in electronic systems using \textit{molecular graphene}~\cite{Freeney2020}, which is constructed by aligning carbon-monoxide molecules regularly on the Cu(111) surface  in terms of the scanning tunneling microscopy~(STM) technique~\cite{Gomes2012,Polini2013}. Such a superstructuring with $C_{6v}$ symmetry can also generate corner modes originating from higher-order topology (HOT)~\cite{Benalcazar2017}, which has been observed in photonic crystals~\cite{Noh2018}, various metamaterials~\cite{Serra-Garcia2018,Peterson2018,Noh2018,Imhof2018,Xue2019,Mittal2019} and electronic artifical lattices~\cite{Kempkes2019}. However, delicate deformations at the angstrom scale in real electronic materials are extremely difficult. While it was proposed that graphene with nanohole arrays~\cite{Kariyado2018} might reduce the difficulty, so far the top-down approach has not suceeded in fabricating nanohole arrays with desired patterns and scales. As the alternative approach, bottom-up synthesis methods have been tried, such as graphene nanoribbon~\cite{Cai2010}, nanoporous graphene~\cite{Moreno2018} and graphyne~\cite{Haley2008,Hu2022} with atomic precision, which yield interesting electronic~\cite{Son2006a,Son2006b,Pan2011,Liu2019} and mechanical properties~\cite{Cranford2011}. 

\begin{figure}[tbp]
      \centering
      \includegraphics{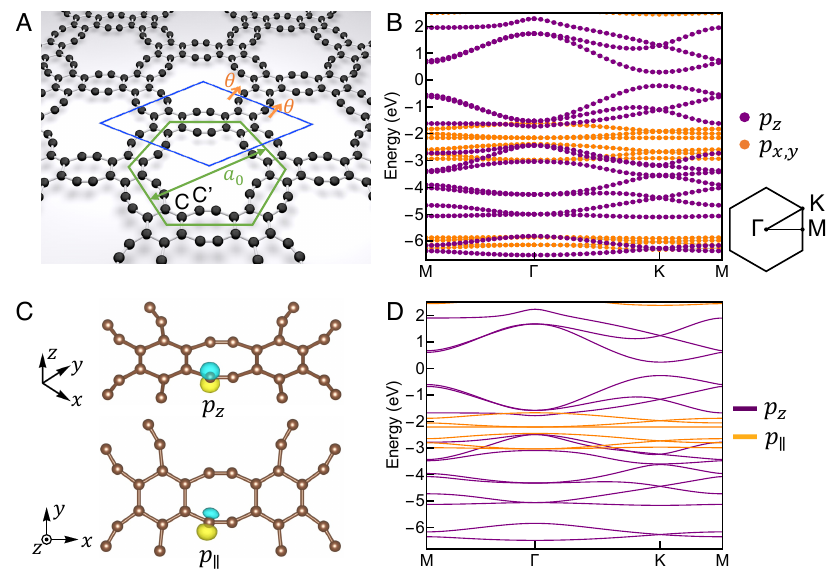}
      \caption{Holey graphyne~(HGY) and its band structures.
        (A)~Stable structure of HGY with lattice constant $a_0$ obtained by DFT calculations. 
        Carbon atoms are divided into two types, C and C', where the former belongs to the benzene rings and the latter only belongs to the octagon.
        The rhombic unit cell corresponds to the precursor in the bottom-up synthesis as discovered in Ref.~\citenum{Liu2022}, whereas the hexagon denotes the highest symmetric unit cell.
        A pair of twist phases $\theta$ are introduced in two $sp$-hybridized bonds for the calculation on the Berry phase $\gamma$ for HOTI. 
        (B)~DFT band structure and the Brillouin zone. The Fermi energy is set to zero, and the eigenstates are projected to $p_z$ and $p_{x,y}$ orbitals. 
        (C)~$p_z$-like and $p_\parallel$-like maximally localized Wannier functions obtained from the localization procedure, where the $p_\parallel$ orbital is the superposition of $p_x$ and $p_y$ orbitals. 
        (D)~Wannier-interpolated band structure obtained from the subspace selected by projecting onto $p_z$ orbitals on each atom and $p_\parallel$ orbitals on atoms  C'.         }
      \label{fig:HGYbulk}
    \end{figure}

  Here, we unveil topological electronic states in the holey graphyne~(HGY) fabricated recently in a bottom-up way~\cite{Liu2022} using first-principles calculations and Wannier TB modeling~\cite{Marzari2012}. HGY is a two-dimensional~(2D) carbon allotrope where benzene rings are connected by two \ch{-C+C\bond{sb}} bonds (Figure~\ref{fig:HGYbulk}A).
  A nontrivial higher-order topological invariant is obtained from the $C_2$ symmetry of the material, and the resultant corner modes appear in nanoflakes with the edge morphology matching to the structure of precursor~\cite{Liu2022}, which are ready for direct experimental observations.
  In addition, we find that intriguing molecular orbitals emerge in the hexagonal unit cell, and specially a band inversion between $g$-like and $h$-like modes gives rise to a nontrivial topology characterized by the $\mathbb{Z}_2$ topological invariant. 
  In order to observe the helical edge states similar to those in QSHE, we propose in-plane hydrogenation in HGY to absorb the $p_{x,y}$ orbitals of carbon atoms in the pristine material, which opens a global topological energy gap at the appropriate energy and makes the helical edge states with opposite orbital angular momenta~(OAM) observable in experiments. 

    By employing density-functional-theory~(DFT) calculations (see Supporting Note S1), we obtain the stable real-space structure of HGY with a lattice constant of $a_0=10.85$~{\AA} (see Supporting Note~S1) and the band structure as depicted in Figures~\ref{fig:HGYbulk}A and \ref{fig:HGYbulk}B, respectively, in good agreement with the results in the previous work~\cite{Liu2022}. The structure in Figure~\ref{fig:HGYbulk}A can be regarded as a network of hexagons (benzene rings) and octagons, and there are two types of carbon sites, one belongs to the hexagon (shared with the octagon) and the other only belongs to the octagon. We name the former and the latter C and C', respectively. 
    For the band structure in Figure~\ref{fig:HGYbulk}B, we project the eigenstates onto the $p_z$ and $p_{x,y}$ orbitals of carbon atoms, and find that occupations are either zero or unity, indicating no hybridization between them due to the mirror symmetry of the 2D material. 
    The $p_z$ orbital is also found orthogonal to the $s$ orbital, and thus no hybridization occurs between these two orbitals. 
    Therefore, the $p_z$ bands can be treated independently of other bands. 

    From the $C_2$ symmetry of the precursor octagon (see Figure~\ref{fig:HGYbulk}A), we surmise a HOT in the HGY.
    In order to explore the possibility, we first employ Wannier localization procedure to obtain the Wannier TB Hamiltonian.
    According to the orbital occupations shown in Figure~\ref{fig:HGYbulk}B, in the Wannierization the eigenstates are projected onto $p_z$ orbitals on each atom, and the resulted maximally localized Wannier function~(MLWF) is illustrated in the upper panel of Figure~\ref{fig:HGYbulk}C (see Supporting Note S1). 
    The Wannier-interpolated band structure for this $p_z$-like MLWF is displayed in Figure~\ref{fig:HGYbulk}D, which reproduces the $p_z$ bands obtained by DFT calculations shown in Figure~\ref{fig:HGYbulk}B.

    With the obtained Wannier TB Hamiltonian, we calculate the Berry phase $\gamma$ for HOT insulator~(HOTI)~\cite{Mizoguchi2019,Kudo2019,Araki2020} associated with a pair of twist phases $\theta$ introduced on the double bonds of HGY as depicted in Figure~\ref{fig:HGYbulk}A (see Supporting Note S1). 
    The Hamiltonian $H(\theta)$ can be separated into two parts as 
      \begin{equation}
        H(\theta)=h_0(\theta)+h_1, \quad h_0(\theta)= -t \sum_{\left<mn\right>} (e^{i\theta}c^{\dagger}_m c_n + e^{-i\theta}c^{\dagger}_n c_m),
      \end{equation}
    where $h_0(\theta)$ with nearest-neighbor hopping energy $t=3.0$~eV obtained by MLWF is for the part with twist phases and $h_1$ is for the rest of the system. Note that since the phase twist is applied only on the selected bonds, this phase twist cannot be gauged out.
    The Berry phase for HOTI is defined as 
      \begin{equation}
        \gamma=-i\int_0^{2\pi}\dd\theta \ev{\partial_\theta}{\Psi(\theta)} \pmod{2\pi}
      \end{equation}
    with $\ket{\Psi(\theta)}$ being the ground state of $H(\theta)$.
    Since the system has the $C_2$ symmetry at the center of octagon, $\gamma$ remains the same when $\theta \rightarrow -\theta$, i.e. $\gamma=-i\int_0^{-2\pi}\dd\theta \ev{\partial_\theta}{\Psi(\theta)} = -i\int_{2\pi}^{0}\dd\theta \ev{\partial_\theta}{\Psi(\theta)} \pmod{2\pi}$, where we shift the integration by $2\pi$ since $H(\theta)=H(\theta+2\pi)$. Knowing that the integration over a close loop $\theta=0 \rightarrow 2\pi \rightarrow 0$ is 0 modulo $2\pi$, $\gamma$ is a $\mathbb{Z}_2$ index being either 0 or $\pi$.
    We find out 
      \begin{equation}
        \gamma=\pi \quad \text{at the $\frac{9}{24}$ filling of $p_z$ bands,}
      \end{equation}
    which corresponds to the filling upto the gap at $-2.5$~eV.
    Our calculation indicates clearly a nontrivial HOT in the HGY.

    The nontrivial HOT will manifest corner modes in flake structures within the topological energy gap of $p_z$ bands, which ranges from $-2.5$~eV to $-1.8$~eV. However, as can be seen in Figure~\ref{fig:HGYbulk}B, this energy gap is covered by the $p_{x,y}$ bands, which hampers the observation of corner modes if the corner modes and the bulk states of $p_{x,y}$ orbitals are energetically close. 
    Therefore, it is necessary to consider a Hamiltonian which includes those $p_{x,y}$ bands around the energy gap. 
    Again we employ Wannier localization procedure, while this time the eigenstates are projected onto a superposition of $p_x$ and $p_y$ orbitals on atoms C'. 
    The resulted MLWF is illustrated in the lower panel of Figure~\ref{fig:HGYbulk}C, which is parallel to the plane of the material and thus named $p_\parallel$. 
    The Wannier-interpolated band structure for this $p_\parallel$-like MLWF, combined with that for the $p_z$-like MLWF, is displayed in Figure~\ref{fig:HGYbulk}D, which matchs perfectly with the DFT band structure shown in Figure~\ref{fig:HGYbulk}B in the energy range of interest.
  
\begin{figure}[t]
        \centering
        \includegraphics{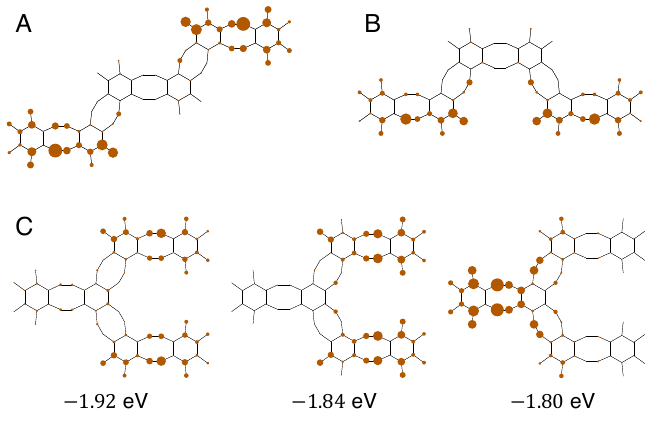}
        \caption{Topological corner modes in nanoflakes of HGY. 
          (A--C) Local density of states (LDOS) $|\psi_i|^2$ for the corner states in all three types of nanoflakes with three rhombic unit cells: (A)~straight type, (B)~boomerang type-I, and (C)~boomerang type-II, with energy $-1.94$~eV, $-1.84$~eV and $-1.80$~eV, respectively.
        }
        \label{fig:HGYflake}
      \end{figure}

    Apparently, in the target bandgap of $p_z$ orbital there are finite DOS of $p_\parallel$. 
    In order to observe topological corner states arising from $p_z$ orbital experimentally, the nanoflake should be small enough to have the bulk spectrum discretized, which makes it possible to distinguish the corner states from bulk states energetically. 
    Here we construct several minimal nanoflakes consisting of three rhombic unit cells with different symmetries which host topological corner states (see other flake structures in Supporting Note~S2). 
    The LDOS $|\psi_i|^2$ for the corner states are visualized in Figure~\ref{fig:HGYflake}, with intensity proportional to size of dots. 
    For the straight type in Figure~\ref{fig:HGYflake}A, the corner state is located at the energy of $-1.85$~eV, with the energy difference of $0.08$~eV to bulk states owing to the finite-size effect (see the energy spectrum in Supporting Note~S2). 
    This energy difference is large enough for the STM technique to detect the corner states by measuring differential conductance~($\dd I/\dd V$) maps. 
    We anticipate that the differential conductance map will show a similar pattern as illustrated in Figure~\ref{fig:HGYflake}A where electrons mainly locate in the two outmost unit cells. 
    From the inversion symmetry of this nanoflake (i.e. $C_2$ symmetry), it is expected that two corner states exist: a bonding state and an antibonding state. 
    Here we only show the antibonding state, since in the bonding state the wave functions at two different corners have a strong overlap with each other, making the bonding state difficult to distinguish from bulk states by only seeing LDOS. 
    For the boomerang type-I in Figure~\ref{fig:HGYflake}B, the corner state is at the energy of $-1.85$~eV, with the energy difference of $0.08$~eV to bulk states. 
    For the boomerang type-II in Figure~\ref{fig:HGYflake}C, the corner states are situated at energies of $-1.92$~eV, $-1.84$~eV and $-1.80$~eV, respectively, and the smallest energy gap between the corner states and bulk states is $0.03$~eV, which can also be observed experimentally. 
    The two corner states with the lower energies possess opposite mirror eigenvalues (see the wave functions in Supporting Note~S2), and their degeneracy is lifted because of the hybridization; while the corner state with the higher energy is a singlet with the wave function localized at the central corner. 

    We also investigate first-order topology which can be characterized by parity index ($C_2$ eigenvalue with respective to the center of hexagonal unit cell shown in Figure~\ref{fig:HGYbulk}A)~\cite{Benalcazar2014}.
    For this purpose, we evaluate the parity index by counting the numbers of parity-even states at both $\Gamma$ and M points~($N^+_{\Gamma/\text{M}}$) for all the valence bands below the global bandgap at $-2.4$~eV and find $(N_{\Gamma}^+, N_\text{M}^+) = (21, 21)$, which apparently suggests that the material is topologically trivial. 
    However, it turns out that the topology of HGY is much richer than it looks at the first glance.

    Since the $p_z$ bands are independent of other bands as discussed in the previous section, we can extract the parity index for the $p_z$ bands which gives $(N_{\Gamma}^+ , N_\text{M}^+) = (3, 5)$. 
    This imbalance of parity index indicates the presence of nontrivial topology. 
    Meanwhile, the parity index for the other bands, namely $s$ and $p_{x,y}$ bands, is $(N_{\Gamma}^+, N_\text{M}^+) = (18, 16)$. 
    By checking the parity index in details, we find out the imbalance originated from the three bands of $p_{x,y}$ orbitals around $E = -3$~eV, which give $(N_{\Gamma}^+, N_\text{M}^+) = (3, 1)$. 
    Therefore, HGY hosts two sets of topological bands, one from the $p_z$ orbital and the other from the $p_{x,y}$ orbitals. 
    Namely, these two orthogonal sets of bands exhibit nontrivial topology, showcasing rich topological characteristics in HGY.
    The Wannier TB model reproduces the parity indices for $p_z$ and $p_{x,y}$ bands obtained by DFT calculations, meaning that it successfuly captures the nontrivial topology of the material.
    
\begin{figure}[tp]
        \centering
        \includegraphics{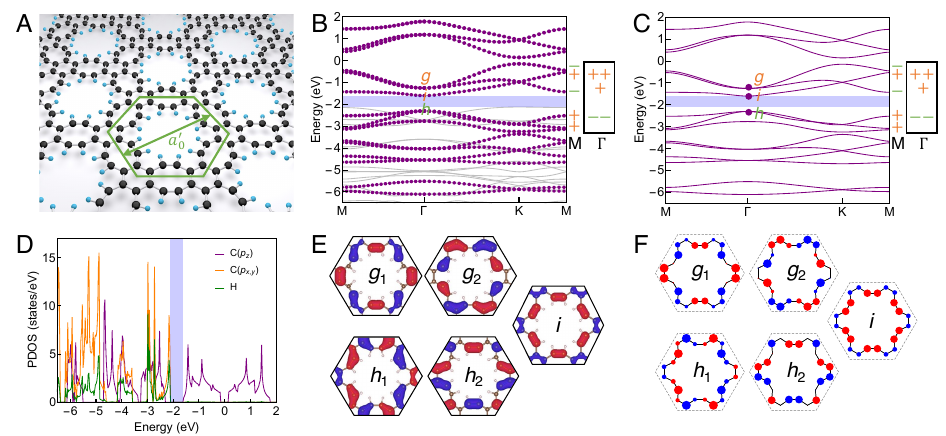}
        \caption{Hydrogenated holey graphyne~(HHGY) and its band structures.
          (A)~Stable structure of HHGY with lattice constant $a'_0$ obtained by DFT calculations, where carbon and hydrogen atoms are colored in black and cyan, respectively. The unit cell with the highest symmetry is denoted by a hexagon. 
          (B)~Band structures obtained by DFT calculations, with eigenstates projected to $p_z$ orbitals and the bandgap shown by a region highlighted in blue. The $g$-, $i$- and $h$-like modes are labeled explicitly, and the parity of the eigenstates at $\Gamma$ and M points for the corresponding five bands are denoted by plus and minus signs. 
          (C)~Same as (B) except for Wannier interpolation which is performed by projecting onto $p_z$ orbitals on each carbon atom.
          (D)~Projected density of states~(PDOS) obtained by DFT calculations, with the bandgap shown by a region highlighted in blue.
          (E)~Wave functions of $g$-, $h$- and $i$-like modes at the $\Gamma$ point obtained by DFT calculations, with red/blue color denotes the plus/minus value. Notation: $g_1\equiv g_{x^4-6x^2y^2+y^4}$, $g_2 \equiv g_{xy(x^2-y^2)}$, $h_1 \equiv h_{x(x^4-10x^2y^2+5y^4)}$, $h_2 \equiv h_{y(5x^4-10x^2y^2+y^4)}$, $i \equiv i_{x^6-15x^4y^2+15x^2y^4-y^6}$. 
          (F)~Same as (E) except for Wannier interpolation, with amplitude represented by size of dots.
        }
        \label{fig:HHGYbulk}
      \end{figure}

    Associated with the imbalance of parity index, we expect that the material manifests topological edge states within the energy gap of $p_z$ bands around $-2$~eV (see Figure~\ref{fig:HGYbulk}B). 
    Unfortunately, as can be seen in Figure~\ref{fig:HGYbulk}D, these edge states are covered in energy by the bulk bands of $p_\parallel$ orbital although orthogonal to each other. 
    In order to solve this issue, we consider in-plane hydrogenation on HGY, where hydrogen atoms are attached to carbon atoms C'.
    The stable structure of hydrogenated HGY~(HHGY) with a lattice constant $a_0'=11.02$~{\AA} obtained by DFT calculations is illustrated in Figure~\ref{fig:HHGYbulk}A (see bond lengths and angles in Supporting Note~S1). 
    The DFT band structure is shown in Figure~\ref{fig:HHGYbulk}B, where a global energy gap of $0.52$~eV is opened at the energy of $-2$~eV compared to the band structure of pristine HGY shown Figure~\ref{fig:HGYbulk}B.
    Similar to the situtation in HGY, the $p_z$ bands in HHGY are orthogonal to other bands because of the mirror symmetry with respect to the horizontal plane.
    Therefore, we perform the Wannierization procedure by projecting the eigenstates onto $p_z$ orbitals on each carbon atom, and the interpolated band structure is shown in Figure~\ref{fig:HHGYbulk}C, reproducing the $p_z$ bands obtained by DFT calculations shown in Figure~\ref{fig:HHGYbulk}B. 

    We display the projected density of states~(PDOS) for HHGY in Figure~\ref{fig:HHGYbulk}D, where the $p_{x,y}$ orbitals of carbon atoms and the $s$ orbital of hydrogen atoms exist below the bandgap at $-2$~eV. 
    This is in sharp contrast to the situation in HGY where the $p_{x,y}$ orbitals fill up almost all the bandgap of $p_z$ orbital as displayed in Figure~\ref{fig:HGYbulk}B. 
    The in-plane hydrogenation leads to the $sp$-hybridization between the $p_{x,y}$ orbitals of carbon atoms and the $s$ orbital of hydrogen atoms, lowering the energy of the original $p_\parallel$ orbital. 
    Therefore, the hydrogenation effectively kills the original $p_\parallel$ orbital and results in an energy gap as large as $0.52$~eV.

    In order to double check our strategy, we re-examine the topology for the $p_z$ bands using parity index, and find imbalance by $(N_{\Gamma}^+, N_\text{M}^+) = (3, 5)$ for the states below the energy gap, indicating the presence of nontrivial topology as expected (see Supporting Note~S3). 
    We can also check the parity index for all states including other orbitals, which gives $(N_{\Gamma}^+, N_\text{M}^+) = (24, 26)$, meaning that the nontrivial topology originates purely from the $p_z$ bands as discussed above. 

    Moreover, in Figures~\ref{fig:HHGYbulk}B and \ref{fig:HHGYbulk}C we notice that the unbalanced parity index is induced by a band inversion between $g$- and $h$-like modes around the bandgap at the $\Gamma$ point. 
    As illustrated in Figures~\ref{fig:HHGYbulk}E and \ref{fig:HHGYbulk}F, the eigenstates $g$-, $h$- and $i$-like modes are named by counting the number of nodes along the perimeter of hexagonal unit cell, i.e., 8, 10, and 12, with even, odd and even parity, respectively. Note that if the green hexagonal cell in Figure~\ref{fig:HHGYbulk}A is isolated and the hoppings between $p_z$ orbitals are uniform, the classification by the number of nodes is exact and the eigenenergy should increase with the number of nodes. However, in Figure~\ref{fig:HHGYbulk}B, $g$-like mode comes above $h$-like mode, signaling a band inversion.
    For the two isolated valence $p_z$ bands with $h$-like mode with odd parity at the $\Gamma$ point and parity-even states at the M point, we evaluate the Wilson loop~\cite{Yu2011,Weng2015} and find a phase winding of $2\pi$, which indicates a nontrivial band topology (see Supporting Note~S4).

    Coming back to the full argument with the actual crystalline symmetry, doubly degenerate $g$- and $h$-like modes are the 2D irreducible representations of the $C_{6v}$ symmetry in the material, while the singlet $i$-like mode is the 1D irreducible representation. 
    With the double degeneracy, we can construct pseudospin states using $g$- and $h$-like modes:
    \begin{equation}
      \ket{g_\pm} = \frac{1}{\sqrt{2}} \left(\ket{g_1} \pm i \ket{g_2}\right), \quad \ket{h_\pm} = \frac{1}{\sqrt{2}} \left(\ket{h_1} \pm i \ket{h_2}\right). 
    \end{equation}
    We regard $\ket{g_+}$ and $\ket{h_+}$ as pseudospin-up states, since their phases of wave functions increase $+8\pi$ and $+10\pi$ counterclockwisely along the perimeter of hexagonal unit cell, which correspond to the states with OAM $+4\hbar$ and $+5\hbar$, respectively, noting that the operator of OAM is $-i\hbar \pdv{\phi}$. The time-reversal counterparts $\ket{g_-}$ and $\ket{h_-}$ are regarded as pseudospin-down states with OAM $-4\hbar$ and $-5\hbar$, respectively. Again, this argument is exact in the case of an isolated cell with a uniform hopping. In this system where there is only $C_{6v}$ symmetry, the eigenstates are classified by OAM upto mod $6\hbar$, and $\pm 4\hbar$ and $\pm 5\hbar$ correspond to $\mp 2\hbar$ and $\mp\hbar$, respectively.

\begin{figure}[tbp]
        \centering
        \includegraphics{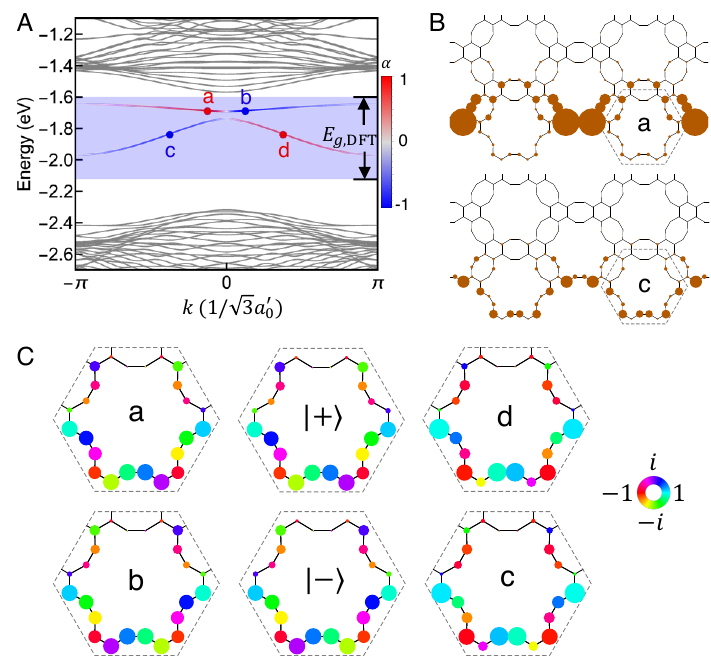}
        \caption{Topological edge states in a ribbon of HHGY.
          (A)~Energy dispersion of a ribbon of HHGY obtained by Wannier TB calculations on $p_z$ orbitals. The region highlighted in blue represents the DFT bulk bandgap, where the edge states are colored by their pseudospin polarizations $\alpha$ (see Supporting Note S5).  
          (B)~LDOS $|\psi_i|^2$ for the edge state ”a” (upper panel) and “c” (lower panel) as labeled in~(A), with the outmost unit cell highlighted by a dashed hexagon. 
          (C)~Comparison between the bulk pseudospin states $\ket{\pm}$ and the wave functions with spinful parts in the outmost unit cell of the four edge states “a” to “d”, with amplitude and phase denoted by size and color of dots, respectively.
        }
        \label{fig:HHGYribbon}
      \end{figure}

    The nontrivial topology characterized by the $\mathbb{Z}_2$ topological index will manifest topological edge states. 
    As can be seen in Figure~\ref{fig:HHGYribbon}A obtained by the Wannier TB calculations on the ribbon structure where the hexagonal unit cells remain intact (see Supporting Note S1), a pair of topological edge states carrying opposite pseudospins appear within the DFT bulk bandgap (see Supporting Notes~S5 for details).
    The LDOS of edge states are shown in Figure~\ref{fig:HHGYribbon}B (for detailed understanding see Supporting Notes~S6), and the wave functions including phases with comparison to the bulk pseudospin states $\ket{\pm}$ ($=(\ket{g_{\pm}} \pm i \ket{h_{\pm}}) / \sqrt{2}$) are shown in Figure~\ref{fig:HHGYribbon}C.
    
    It should be noted that although the edge morphology exhibiting the topological edge states in this work cannot be achieved by the available precursor directly, and thus trials on edge treatments are required, such as out-of-plane hydrogenation.

  In summary, we uncover topological electronic states in holey graphyne, through DFT calculations and Wannier TB modeling. 
  Firstly, a nontrivial $\mathbb{Z}_2$ higher-order topological invariant is obtained arising from the $C_2$ symmetry of the material. We display explicitly the minimal nanoflakes hosting the corresponding corner states, which can be fabricated using the same precursor in the experiment~\cite{Liu2022}, and the corner modes are ready to be observed by measuring differential conductance maps using STM.
  While so far the observation of topological corner modes in electronic systems has been limited to artificial lattices made by positioning carbon-monoxide molecules on Cu(111) surface~\cite{Kempkes2019}, holey graphyne could be the non-artificial material where topological electronic corner modes are experimentally observed.
  Secondly, by introducing an in-plane hydrogenation, holey graphyne opens a large energy gap of 0.52~eV at $-2$~eV, with a nontrivial topology arising from the band inversion between $g$- and $h$-like modes and characterized by a $\mathbb{Z}_2$ invariant.
  Topological edge states carrying opposite OAMs popagate in opposite directions, indicating potential applications in orbitronic devices with low energy loss workable at room temperature.
  Our work points out that holey graphyne serves as an ideal platform to observe topological electronic corner modes and edge states with OAMs protected by a large energy gap, opening a new avenue for further investigation and exploitation of topology in 2D carbon allotropes for both scientific understanding and pratical applications.

\begin{suppinfo}
  The Supporting Information is available free of charge at (website) 

  \begin{quotation}\noindent
    Section~S1 on first-principles calculations and Wannier localization procedures;
    Section~S2 on energy spectra and corner states of various HGY nanoflakes;
    Section~S3 on adiabatic deformation of Wannier TB models between HGY and HHGY;
    Section~S4 on Wilson loop calculations on HHGY;
    Section~S5 on weights of spinful and spinless parts for the wave functions of edge states in the HHGY ribbon;
    Section~S6 on analysis on the wave functions of edge states in the HHGY ribbon
    (PDF)
  \end{quotation}

\end{suppinfo}

\section*{Declaration of interests} 
The authors declare no competing financial interests.

\begin{acknowledgement}
  This work is supported by CREST, JST (Core Research for Evolutionary Science and Technology, Japan Science and Technology Agency) (Grant Number JPMJCR18T4).
\end{acknowledgement}

%\bibliography{TopoHGY}

\providecommand{\latin}[1]{#1}
\makeatletter
\providecommand{\doi}
  {\begingroup\let\do\@makeother\dospecials
  \catcode`\{=1 \catcode`\}=2 \doi@aux}
\providecommand{\doi@aux}[1]{\endgroup\texttt{#1}}
\makeatother
\providecommand*\mcitethebibliography{\thebibliography}
\csname @ifundefined\endcsname{endmcitethebibliography}
  {\let\endmcitethebibliography\endthebibliography}{}

%%%%%%%%%%%%%%%%%%%%%%%%%%%%%%%%%%%%%%%%%%%%%%%%%%%%%%%%%%%
%%%%%%%%%% Merge with supplemental information %%%%%%%%%%
\pagebreak
%\widetext
\begin{center}
\textbf{\Large Supporting Information for \\
Topological electronic states in holey graphyne}
\end{center}

%---------------------------------------------
\setcounter{equation}{0}
\setcounter{figure}{0}
\setcounter{page}{1}
\renewcommand{\theequation}{S\arabic{equation}}
\renewcommand{\thefigure}{S\arabic{figure}}
\renewcommand{\bibnumfmt}[1]{(S#1)}
\renewcommand{\citenumfont}[1]{S#1}
%--------------------------------------------

\section{First-principles calculations and Wannier localization procedures}
  
  First-principles calculations are performed within the DFT scheme using the Vienna \textit{Ab initio} Simulation Package~\cite{Kresse1996}, where the projector augmented-wave method~\cite{Kresse1999}, the Perdew-Becke-Erzenhof type generalized gradient approximation~\cite{Perdew1996} for the exchange-correlation potential and a plane-wave basis set with a cutoff energy of 520~eV are adopted. A $11\times11\times1$ $k$-point mesh is used for both structure relaxations and self-consistent calculations. The structure relxations are performed until the Hellmann-Feynman forces acting on ions are smaller than $10^{-4}$~eV/{\AA} and the energy tolerances are below $10^{-6}$~eV/atom. Graphyne sheets are separated by a vacuum layer of 2~nm, ensuring that interlayer couplings are negligible. The data post-processing is done using VASPKIT~\cite{Wang2021}.

\begin{figure}[tb]
    \centering
    \includegraphics[width=\textwidth]{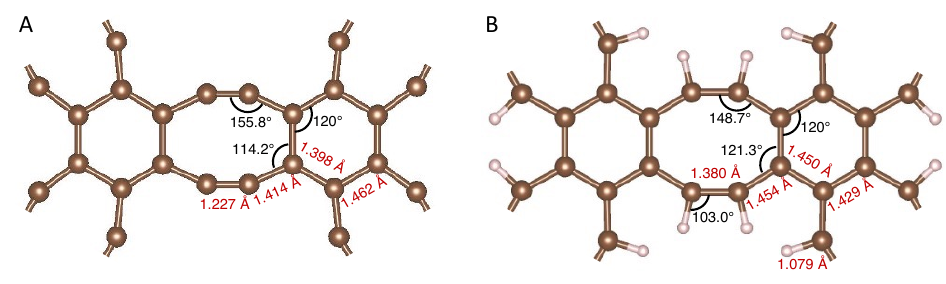}
    \caption{Bonding information.
      (A) Bond lengths and angles of HGY obtained by DFT calculations. 
      (B) Same as (A) except for HHGY. 
    }
    \label{fig:bond}
    \end{figure}

  In Figure~\ref{fig:bond}A we give the bond lengths and angles obtained by the ion relaxation in DFT calculations for HGY, which successfully reproduces the bonding information in Ref.~\citenum{Liu2022}. With the $C_{6v}$ symmetry of HGY, the shown information fixes the crystalline structure completely, namely we can derive all other bond lengths and angles. The lattice constant is $a_0=10.85$~{\AA}, which has already been given in the main text.
  For HHGY, the bond lengths and angles are given in Figure~\ref{fig:bond}B, with the lattice constant $a'_0=11.02$~{\AA}.

  MLWF are obtained from data in first-principles calculations using WANNIER90~\cite{Pizzi2020}. For HGY, the Wannier localization procedures are performed by projecting the eigenstates onto $p_z$ orbitals on each atom and $p_\parallel$ orbitals on atoms C'. The outer energy window is set as $E_{\text{out}}\in[-7~\text{eV},12~\text{eV}]$ to include all $p$ orbitals, and the frozen energy window is set as $E_{\text{froz}}\in[-5.3~\text{eV},2.4~\text{eV}]$.

  For HHGY, the projection orbitals are $p_z$ orbitals on each carbon atom and bonding $sp$ orbitals at the centers of C--H bonds. The outer energy window is the same as the case for HGY, while the frozen energy window starts from $-4.5$~eV to avoid other orbitals. We show only the interpolated band structure for $p_z$ orbitals in Figure~3C in the main text, since the $sp$ orbitals are orthogonal to the $p_z$ orbitals and irrelevant for our discussion.

  In the calculation of the HOT index $\gamma$ for HGY, a pair of twist phases $\theta$ are introduced in a supercell consisting of $3\times3$ rhombic unit cells with periodic boundary condictions applied in two in-plane directions.
  For the ribbon structure of HHGY, the calculation is performed on a supercell composed of 20 hexagonal unit cells using a Wannier TB Hamiltonian.

\section{Energy spectra and corner states of various HGY nanoflakes}

\begin{figure}[tb]
      \centering
      \includegraphics[width=\textwidth]{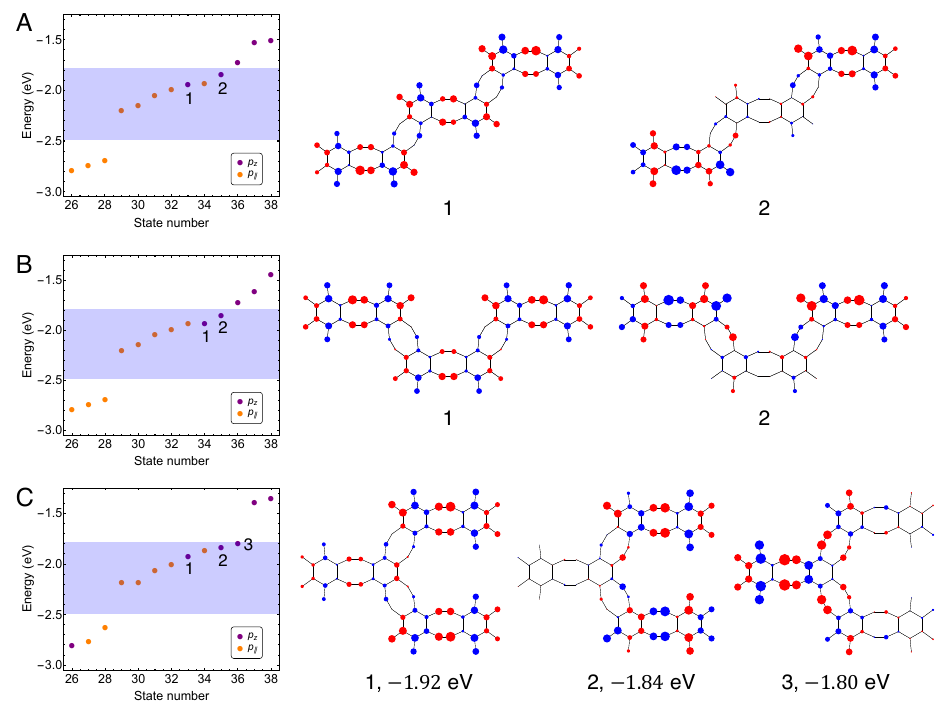}
      \caption{Energy spectra and corner modes of HGY nanoflakes with three rhombic unit cells.
        (A) (left panel) Energy spectrum for straight type with corner states labeled by numbers. The region highlighted in blue represents the bulk bandgap of $p_z$ orbitals shown in Figure~3C in the main text. 
        (right panel) Wave functions of corner modes with amplitude represented by dot size and positive/negative signs in red/blue. 
        (B and C) Same as (A) except for boomerang type-I (B) and type-II (C). 
      }
      \label{fig:flakeUC3}
    \end{figure}

  For the nanoflakes with one and two rhombic unit cells, the bulk and corner are not well-defined, and thus we start with nanoflakes with three rhombic unit cells, where there are three different configurations as shown in Figure~\ref{fig:flakeUC3}.
  In Figure~\ref{fig:flakeUC3}A we display the energy spectrum for the straight type, and find two in-gap states of $p_z$ orbital. These two in-gap states possess even and odd parity with respect to the center of nanoflake, which correspond to the bonding and antibonding states, respectively, as shown in \Cref{fig:flakeUC3}A. 
  Although these two states originate from the nontrivial HOT in this nanoflake with two corners, only the antibonding state can be regarded as a corner state, while it is not the case for the bonding state which has strong overlap at the central unit cell. 
  The energy of the antibonding state is $-1.85$~eV, and the energetically closest state is the bulk state of $p_\parallel$ orbital at $-1.93$~eV. The energy difference of $0.08$~eV is enough for STM technique to detect the antibonding state. 
  Therefore, the nanoflake with three rhombic unit cells is the minimal one hosting topological corner modes which can be detected experimentally.

  For the boomerang type-I shown in Figure~\ref{fig:flakeUC3}B, similar to the straight type discussed in the previous paragraph, there are bonding (mirror-even) and antibonding (mirror-odd) states in the energy gap, and only the latter one can be regarded as a corner state (here, the mirror plane is vertical in the figure).
  The energy of the antibonding state is $-1.85$~eV, and the energetically closest state is the bonding state at $-1.93$~eV. 
  The energy difference of $0.08$~eV is enough for STM technique to detect this antibonding state.

  For the boomerang type-II shown in Figure~\ref{fig:flakeUC3}C, different from the two previous cases, there are three in-gap states because this nanoflake has three corners. 
  States 1 and 2 with energy of $-1.92$~eV and $-1.84$~eV respectively are the bonding (mirror-even) and antibonding (mirror-odd) states localized at the upper and lower corners, while state 3 with energy of $-1.80$~eV is a mirror-even state localized at the centeral corner (here, the mirror plane is horizontal in the figure).
  All these three in-gap states can be regarded as corner states from their wave functions (see LDOS in Figure~2 in the main text for a clearer visualization).
  The energetically closest bulk state is at $-1.87$~eV, meaning that the smallest energy gap between the corner states and bulk states is $0.03$~eV, still large enough for STM technique to detect all the three corner states.

    \begin{figure}[tb]
      \centering
      \includegraphics[width=\textwidth]{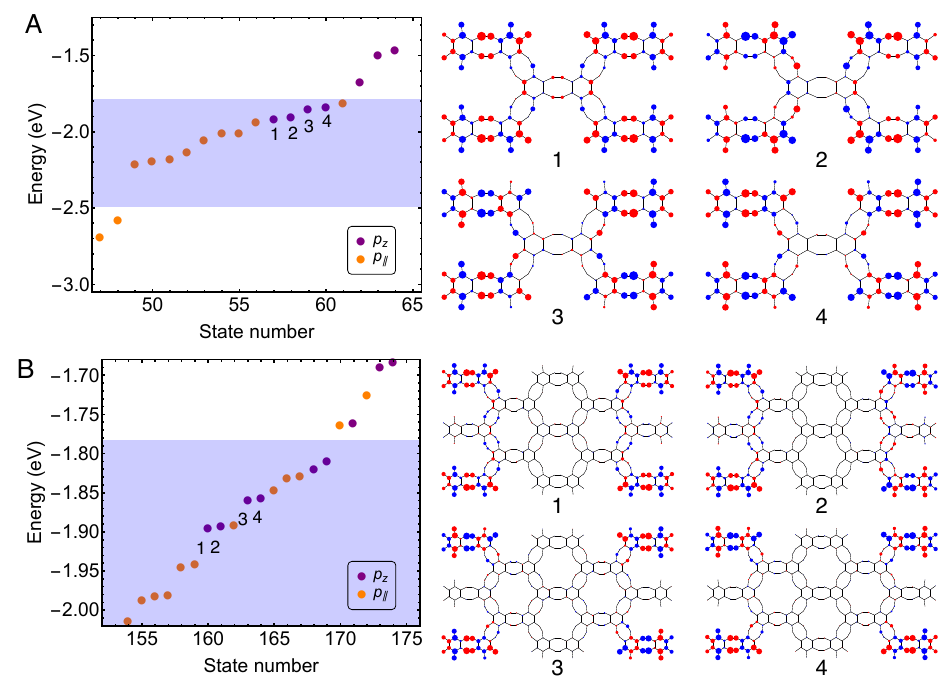}
      \caption{Energy spectra and corner modes of larger HGY nanoflakes.
        (A) (left panel) Energy spectrum for the nanoflake with five rhombic unit cells. The corner states are labeled by numbers, and the region highlighted in blue represents the bulk bandgap of $p_z$ orbitals shown in Figure~3C in the main text. 
        (right panel) Wave functions of corner modes with amplitude represented by dot size and positive/negative signs in red/blue. 
        (B) Same as (A) except for a nanoflake with 13 unit cells.
      }
      \label{fig:flakeUCmore}
    \end{figure}

  In order to show that the in-gap states are second-order topological corner states instead of first-order topological end states, we perform calculations on nanoflakes with larger sizes.
  In Figure~\ref{fig:flakeUCmore}A we show a nanoflake with five rhombic unit cells, which has the highest symmetry, namely, mirror symmetry along both horizontal and vertical directions.
  Four corner states, $s$, $p_x$, $d_{xy}$ and $p_y$ modes, appear within the energy gap, corresponding to state 1 to 4 in Figure~\ref{fig:flakeUCmore}A respectively, with slightly different energies due to the coupling between corner states. 
  Increasing the number of unit cells to 13, it is clear that the four states in the nanoflake demonstrate strong localization at the corners within the outmost unit cells, not along the edge, as shown in Figure~\ref{fig:flakeUCmore}B. 
  It is not difficult to imagine that, when the flake is large enough so that the coupling between two corners becomes negligible, the two mirror symmetries guarantee four-fold degenerate states with each localized at one of the four corners.
  Comparing the wave functions at the upper-right corner in all the corner states shown in Figures~\ref{fig:flakeUC3} and \ref{fig:flakeUCmore} (except for state 3 in Figure~\ref{fig:flakeUC3}C where the wave function localizes at the left corner), we can see that they all look similar, indicating that they originate from the same HOT. 
  Therefore, the nanoflakes with three rhombic unit cells are the minimal ones which host topological corner states detectable experimentally.

  In principle, $p_\parallel$ orbitals could also host HOTI states. However, the corresponding energy gaps are too small for experimental observations.

\section{Adiabatic deformation of Wannier TB models between HGY and HHGY}

\begin{figure}[tb]
      \centering
      \includegraphics[width=\textwidth]{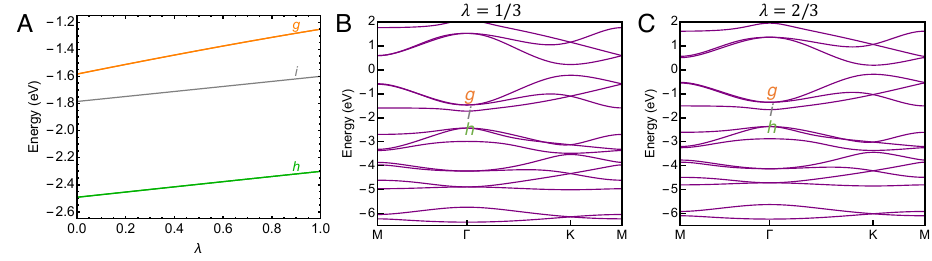}
      \caption{Adiabatic deformation of Wannier TB models between HGY and HHGY.
        (A) $\lambda$ dependence of energies of $g$-, $h$- and $i$-like modes in adiabatic deformation on hoppings with $\lambda$ defined in \Cref{eq:hamLambda}. No energy crossing occurs in these three modes. 
        (B and C) Band structures with hoppings given by $\lambda=1/3$ and $\lambda=2/3$. $h$-like modes are at the top of the valence bands and $i$-like mode is at the bottom of the conduction band.
      }
      \label{fig:adiabatic}
    \end{figure}

  In the main text, we have seen that HGY and HHGY share the same parity index $(N^+_\Gamma,N^+_\text{M})$ for the $p_z$ bands. In order to confirm that these two phases are the same in topology, we consider a continuous transformation between the effective models for HGY and HHGY. For this purpose, the TB models derived using the MLWF method are utilized, 
    \begin{equation}\label{eq:hamLambda}
      H_{\lambda}(\bm{k})=(1-\lambda)H^{\text{TB}}_{\text{HGY}}(\bm{k})+\lambda H^{\text{TB}}_{\text{HHGY}}(\bm{k}),
    \end{equation}
  where $H^{\text{TB}}_{\text{HGY}}(\bm{k})$ and $H^{\text{TB}}_{\text{HHGY}}(\bm{k})$ are the TB Hamiltonians for HGY and HHGY, respectively. This interpolation is possible since $H^{\text{TB}}_{\text{HGY}}(\bm{k})$ and $H^{\text{TB}}_{\text{HHGY}}(\bm{k})$ have the same size as far as the $p_z$ bands are concerned. The band edge energies (at the $\Gamma$ point) as a function of $\lambda$ are shown in Figure~\ref{fig:adiabatic}A, where there is no band crossing. Therefore, the electronic structures for HGY and HHGY are adiabatically connected without gap closing. As can be read from Figures~\ref{fig:adiabatic}B and \ref{fig:adiabatic}C\, the electronic structures are also smoothly transformed from HGY to HHGY, preserving the target energy gap.
  
\section{Wilson loop calculations on HHGY}

\begin{figure}[tb]
      \centering
      \includegraphics[width=\textwidth]{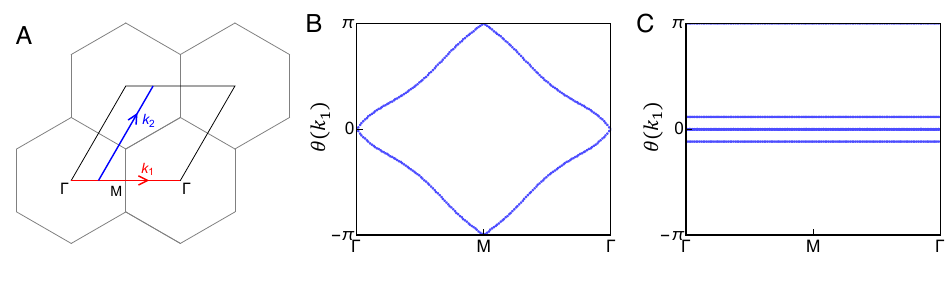}
      \caption{Wilson loop of HHGY.
        (A) Paths for calculations on Wilson loop.
        (B) Wilson loop spectrum along $k_1$ direction for the two isolated bands just below the energy gap at -2~eV shown in Figure~3C in the main text, which demonstrates a nontrivial $\mathbb{Z}_2$ topology. 
        (C) Same as (B) except for all nine valence bands.
      }
      \label{fig:WL}
    \end{figure}

  The Wilson loop along the $k_1$ direction in the BZ in \Cref{fig:WL}A\ is given by~\cite{Yu2011,Weng2015}
  \begin{equation}
    \mathcal{W}(k_1) = \mathcal{P} \exp \left[ -i \oint_{k_2} \dd k_1 \mathcal{A}_1 (k_1,k_2) \right],
  \end{equation}
  where the path-ordered integration is at fixed $k_1$ with $k_2$ across the BZ (for example from $k_2=0$ to $k_2=2\pi$ as shown in \Cref{fig:WL}A), and $\mathcal{A}_1 (k_1,k_2)$ is the Berry connection matrix along $k_1$ direction with the element defined as
  \begin{equation}
    \mathcal{A}_1^{mn} (k_1,k_2) = \mel{u_m(k_1,k_2)}{\partial_{k_1}}{u_n(k_1,k_2)}
  \end{equation}
  with $\ket{u_m(k_1,k_2)}$ the eigenstate in the $m$-th band. The phase of eigenvalue of Wilson loop is 
  \begin{equation}
    \theta(k_1) = \Im \left[ \ln \text{eig}\left( \mathcal{W}(k_1) \right) \right].
  \end{equation}

  We calculate the phase of Wilson loop eigenvalue for the two isolated bands just below the energy gap at $-2$~eV in HHGY (see Figure~3C in the main text), where the $\Gamma$ point is occupied by two degenerate $h$-like modes with odd parity whereas the M point is occupied by two states with even parity. 
  As displayed in \Cref{fig:WL}, taking the path $k_1=0 \rightarrow \pi \rightarrow 2\pi$, i.e., $\Gamma\rightarrow\text{M}\rightarrow\Gamma$, $\theta(k_1)$ has a phase winding of $2\pi$, indicating the nontrivial band topology. 
  With all nine valence bands taken into calculation, there is no phase winding in $\theta(k_1)$ as shown in \Cref{fig:WL}C, indicating a trivial topology.
  The phase winding situations in Figures~\ref{fig:WL}B and \ref{fig:WL}C signal a fragile topology in the top two valence bands~\cite{Cano2018,Po2018}.

\section{Weights of spinful and spinless parts for the wave functions of edge states in the HHGY ribbon}

\begin{figure}[tbp]
      \centering
      \includegraphics[width=\textwidth]{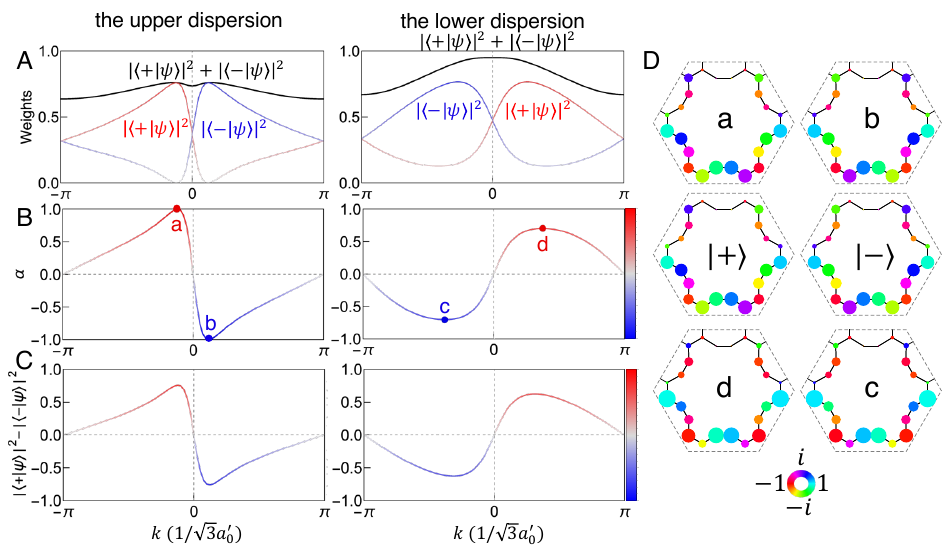}
      \caption{Weights of spinful parts for the wave functions of edge states in the ribbon of HHGY in Figure~4 in the main text.
        (A) Weights of pseudospin-up and -down parts for the wave functions of edge states in the upper and lower dispersions shown in Figure~4A in the main text. 
        (B) Same as (A) except for $\alpha$.
        (C) Same as (A) except for the weight difference between pseudospin-up and -down parts.
        (D) Reproduced from Figure~4C in the main text for convenience.
      }
      \label{fig:polarization}
    \end{figure}

  For the wave functions of edge states, we compare the weights between spinful parts from $g$- and $h$-like modes and the spinless part from the $i$-like mode, and find that the spinful parts predominate in these edge states ($74{\sim}95\%$ around $k=0$, see Figures~\ref{fig:polarization}A). 
  Therefore, we first focus on the spinful parts, and introduce a quantity to measure the polarization of pseudospin:
    \begin{equation}\label{eq:polarization}
      \alpha = \frac{|\braket{+}{\psi}|^2 - |\braket{-}{\psi}|^2}{|\braket{+}{\psi}|^2 + |\braket{-}{\psi}|^2},
    \end{equation}
  where $\ket{\psi}$ is the wave function in the outmost unit cell of ribbon, and $\ket{\pm} = (\ket{g_{\pm}} \pm i \ket{h_{\pm}}) / \sqrt{2}$ denoting the pseudospin-up/-down state in the hexagonal unit cell (see Refs.~\citenum{Wu2015} and \citenum{Kariyado2018} for the analytical solutions of edge states $\ket{p_{\pm}} \pm i \ket{d_{\pm}}$, where a $p$-$d$ band inversion occurs).

  In Figures~\ref{fig:polarization}B, we display the pseudospin polarization $\alpha$ for the edge states as a function of Bloch momentum. 
  In the upper energy dispersion, $\alpha$ remains positive for $k<0$ while negative for $k>0$, showing that the dispersion is divided into two parts, with the left/right one occupied by pseudospin-up/-down states. 
  For edge state ``a'', $\alpha$ reaches nearly unity at $k=-\pi / 8$, indicating a well developed pseudospin-up characteristic, whereas edge state ``b'' at $k=\pi/8$, as the time-reversal counterpart for state ``a'', exhibits a well developed pseudospin-down characteristic.
  Due to the absence of chiral symmetry, the degeneracy at $k=0$ is lifted with a small energy gap, as compared with a large energy gap at $k=\pm\pi$, which leads to the two states $\ket{g_1}-\ket{h_2}$ (mirror-even) and $\ket{g_2}+\ket{h_1}$ (mirror-odd) with zero spin polarization at $k=0$ (see Figure~3E in the main text and Figure~\ref{fig:EdgeState}A).
  In the simpler modulated honeycomb lattice model where the topological nontriviality is manifested by the band inversion between $p$- and $d$-modes, the chiral symmetry (sublattice symmetry) combined with the mirror symmetry is the key to suppress the minigap~\cite{Kariyado2017}, in contrast to the present system where the chiral symmetry is not active.

  In the lower energy dispersion, the edge states for $k<0$ ($k>0$) are pseudospin-down (-up), which is opposite to the upper energy dispersion. Therefore, as depicted in Figure~4A in the main text, the edge states with opposite pseudospins counterpropagate on the same edge of ribbon, exhibiting the feature of pseudospin-momentum locking, reminiscent of QSHE~\cite{Bernevig2006}.

  In Figure~4B in the main text, we plot the LDOS $|\psi_i|^2$ for edge state ``a'' and ``c'', which demonstrate a strong localization on the edge. 
  The LDOS for edge states ``b'' and ``d'' are identical to those for ``a'' and ``c'', respectively, as guaranteed by TRS. 
  We extract the wave functions with only spinful parts in the outmost unit cell, which is illustrated in Figure~\ref{fig:polarization}D, and compare them with the pseudospin-up state $\ket{+}$ and the pseudospin-down state $\ket{-}$. 
  The wave functions for ``a'' and ``b'' show a perfect match to $\ket{+}$ and $\ket{-}$, respectively, which is consistent with the almost full pseudospin polarization. 
  For wave functions ``c'' and ``d'', there are slight deviations in amplitudes and phases, indicating a small mixing between the pseudospin-up and -down states. The time-reversal symmetry between states ``a'' and ``b'' (as well as ``c'' and ``d'') is reflected in the opposite phases of their wave functions.

  In Figure~\ref{fig:polarization}A, we display the weights of spinful parts for the wave functions of edge states in the HHGY ribbon shown in Figure~4 in the main text.
  For the wave functions of edge states in the upper dispersion, as shown in the left panel of \Cref{fig:polarization}A, the weight of spinful parts $|\braket{+}{\psi}|^2+|\braket{-}{\psi}|^2$ is $74\%$ at $k=0$ and has a maximum $76\%$ at $k=0.15\pi$. The rest is the spinless part, which comes from the $i$-like band. 
  Although the edge states originate from the $g$-$h$ band inversion, there is a moderate contribution from the $i$-like band, since it is the lowest conduction band around the $\Gamma$ point lying between the $g$- and $h$-like bands and is close to the upper edge dispersion energetically (see Figures~3B and 4A in the main text).
  For the wave functions of edge states in the lower dispersion, as shown in the right panel of  \Cref{fig:polarization}A, the weight of spinful parts is $90\%\sim95\%$ within $k=\pm0.3\pi$.
  For the wave functions in both the upper and lower edge dispersions, the spinful parts dominate around $k=0$, which justifies the treatment in the main text where the spinful parts are focused.

  In Equation~\ref{eq:polarization} $\alpha$ is defined as the pseudospin polarization within spinful parts of wave function. When the spinless part of wave function is included, the polarization is the weight difference between pseudospin-up and -down parts of wave function. The polarizations for the upper and lower edge dispersions in Figure~\ref{fig:polarization}C have lower peak values of 0.76 and 0.62 as compared with 1.00 and 0.70 in $\alpha$ in Figure~\ref{fig:polarization}B, respectively. The difference is larger in the upper dispersion, because it is energetically closer to the $i$-like band.

\section{Analysis on the wave functions of edge states in the HHGY ribbon}

    \begin{figure}[tb]
      \centering
      \includegraphics[width=\textwidth]{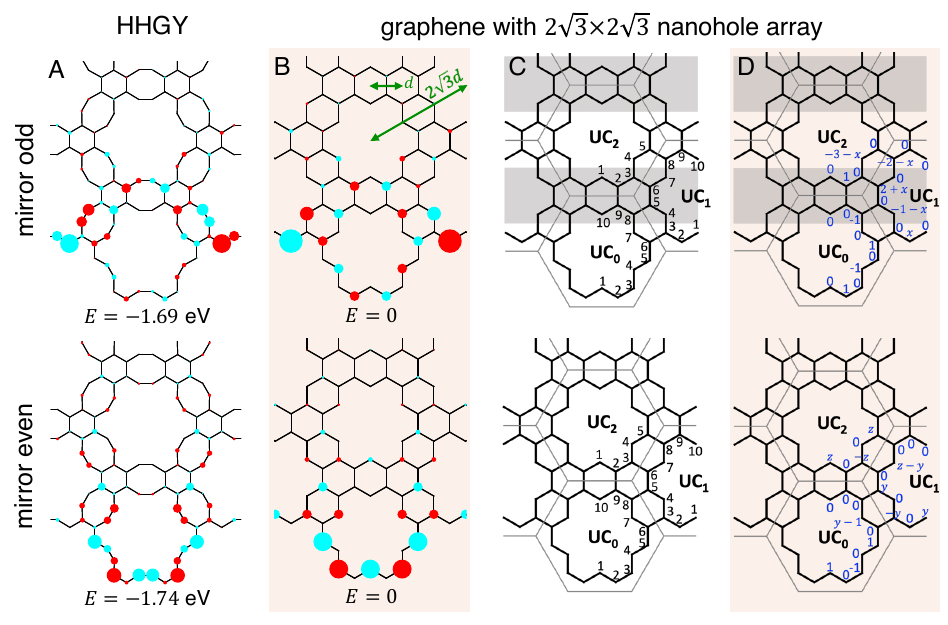}
      \caption{Solutions for edge states.
        (A) Wave functions of edge states in the HHGY ribbon with energies of $-1.69$~eV and $-1.74$~eV at $k=0$ shown in Figure~4C in the main text obtained numerically using the Wannier TB model. Their mirror symmetries are odd and even, respectively. 
        (B) Analytical solutions for zero-energy modes at $k=0$ in a ribbon of graphene with $2\sqrt{3}\times2\sqrt{3}$ nanohole array, where the lattice constant in bulk is $2\sqrt{3}d$ with $d$ being the lattice constant of pristine graphene. 
        (C) Schematic figures labeling the sites in three outmost unit cells on the edge, with the upper/lower one for the mirror-odd/even state. 
        (D) Same as (D) except for analytical solutions of zero-energy modes. The values of $x$, $y$ and $z$ are derived in the text.
      }
      \label{fig:EdgeState}
    \end{figure}

  The edge wave functions corresponding to the LDOS in Figure~4C in the main text is shown in \Cref{fig:EdgeState}A, where state ``a'' dominates on the concave region of edge, while state ``c'' dominates on the outmost carbon chain. 
  The system can be modeled by a simple TB model with nearest-neighbor (NN) hopping called graphene with $2\sqrt{3}\times2\sqrt{3}$ nanohole array as shown in \Cref{fig:EdgeState}B (see Figure~1F in Ref.~\citenum{Kariyado2018} for graphene with $4\sqrt{3}\times4\sqrt{3}$ nanohole array), presuming that the NN hopping between two atoms C' in HHGY shown in Figure~3A in the main text is large enough to treat them as a composite atom.
  The TB Hamiltonian is given by~\cite{Kariyado2018}
    \begin{equation}\label{eq:TB}
       H = -t\sum_{\left<mn\right>}c_m^{\dagger} c_n, 
    \end{equation}
  where only the NN hopping $t=-2.7$~eV is considered.
  The eigenstate of this Hamiltonian can be expressed using $\ket{\psi_{0,1},\psi_{0,2},\cdots,\psi_{i,j},\cdots,\psi_{N,18}}$, where $\psi_{i,j}$ is the local wave function at site $j$ in the $i$-th hexagonal unit cell.
  From the hexagonal unit cell shown in \Cref{fig:EdgeState}C, we can see that this system has chiral (sublattice) symmetry.
  With the Hamiltonian in \Cref{eq:TB}, one can obtain the zero-energy edge modes at $k=0$ in a ribbon structure shown in \Cref{fig:EdgeState}B analytically with the help of the chiral symmetry and the mirror symmetry, as shown in what follows.

  When the state at $k=0$ is concerned, the ribbon can be treated as a cylinder with the diameter to be the lattice constant of supercell~\cite{Kariyado2017}.
  Take the upper panel of \Cref{fig:EdgeState}C for example, the left and right sides are connected with each other, and the edge is of open boundary condition (the top edge is not shown in the figure). 
  In order to obtain the zero-energy mode with odd mirror symmetry, we use the upper panel of \Cref{fig:EdgeState}C as the schematic figure for derivation of wave functions.
  With the mirror symmetry, we only need to consider the wave functions at ten sites in one unit cell.
  When the mirror symmetry is odd, the wave functions at site 1 and 10 are zero, 
  \begin{equation}
    \psi^{\text{o}}_{i,1}=\psi^{\text{o}}_{i,10}=0, \quad i\in N,
  \end{equation}
  where the superscript ``o'' denotes odd mirror symmetry and $i$ labels the unit cell.
  Setting $\psi^{\text{o}}_{0,2}=1$ and considering the state is of zero energy, referring to site 2 in the 0th unit cell there is a relation: $\psi^{\text{o}}_{0,1}+\psi^{\text{o}}_{0,3}=0$, giving
  \begin{equation}
    (\text{site}~0,2)\quad \psi^{\text{o}}_{0,3} = -\psi^{\text{o}}_{0,1} = 0.
  \end{equation}
  Similarly, we can obtain the relations for the following sites:
  \begin{equation}
    \begin{split}
      & (\text{site}~0,3)\quad \psi^{\text{o}}_{0,4} = -\psi^{\text{o}}_{0,2} = -1, \\
      & (\text{site}~0,4)\quad \psi^{\text{o}}_{0,5} = -\psi^{\text{o}}_{0,3} = 0, \\
      & (\text{site}~0,5)\quad \psi^{\text{o}}_{0,6} = -\psi^{\text{o}}_{0,4} = 1, \\
      & (\text{site}~1,2)\quad \psi^{\text{o}}_{1,3} = -\psi^{\text{o}}_{1,1} = 0, \\
      & (\text{site}~0,6)\quad \psi^{\text{o}}_{0,7} = -\psi^{\text{o}}_{0,5}-\psi^{\text{o}}_{1,3} = 0. \\
    \end{split}
  \end{equation}
  Set $\psi^{\text{o}}_{1,2}=x$ and the relations of wave functions at the following sites can be derived:
  \begin{equation}\label{eq:oddwfuc1}
    \begin{split}
      & (\text{site}~1,3)\quad \psi^{\text{o}}_{1,4} = -\psi^{\text{o}}_{0,6}-\psi^{\text{o}}_{1,2} = -1-x, \\
      & (\text{site}~0,7)\quad \psi^{\text{o}}_{0,8} = -\psi^{\text{o}}_{0,6} = -1, \\
      & (\text{site}~1,4)\quad \psi^{\text{o}}_{1,5} = -\psi^{\text{o}}_{1,3} = 0, \\
      & (\text{site}~0,8)\quad \psi^{\text{o}}_{0,9} = -\psi^{\text{o}}_{0,7}-\psi^{\text{o}}_{1,5} = 0, \\
      & (\text{site}~1,5)\quad \psi^{\text{o}}_{1,6} = -\psi^{\text{o}}_{0,8}-\psi^{\text{o}}_{1,4} = 2+x, \\
      & (\text{site}~0,9)\quad \psi^{\text{o}}_{2,2} = -\psi^{\text{o}}_{0,8}-\psi^{\text{o}}_{0,10} = 1, \\
      & (\text{site}~2,2)\quad \psi^{\text{o}}_{2,3} = -\psi^{\text{o}}_{0,9}-\psi^{\text{o}}_{2,1} = 0, \\
      & (\text{site}~1,6)\quad \psi^{\text{o}}_{1,7} = -\psi^{\text{o}}_{1,5}-\psi^{\text{o}}_{2,3} = 0, \\
    \end{split}
  \end{equation}
  which are the wave functions in the gray region acrossing the 1st unit cell (see the upper panel of \Cref{fig:EdgeState}C).
  Stepping further into the bulk, the wave functions in the white stripe region acrossing the 2nd unit cell should have similar relations as \Cref{eq:oddwfuc1}, since the gray region becomes the white region by shifting a unit vector considering the periodicity of the model away from the edge. 
  Therefore, the wave functions for the stripe regions acrossing the $i$-th unit cell are:
  \begin{subequations}\label{eq:oddwf}
    \begin{align}
      & \psi^{\text{o}}_{i,4} = -\psi^{\text{o}}_{i-1,6}-\psi^{\text{o}}_{i,2}, \label{eq:oddwf1} \\
      & \psi^{\text{o}}_{i-1,8} = -\psi^{\text{o}}_{i-1,6}, \label{eq:oddwf2} \\
      & \psi^{\text{o}}_{i,6} = -\psi^{\text{o}}_{i-1,8}-\psi^{\text{o}}_{i,4}, \label{eq:oddwf3} \\
      & \psi^{\text{o}}_{i+1,2} = -\psi^{\text{o}}_{i-1,8}-\psi^{\text{o}}_{i-1,10}=-\psi^{\text{o}}_{i-1,8}, \label{eq:oddwf4}
    \end{align}
  \end{subequations} 
  with $i\in N_+$, and the following wave functions are zero:
  \begin{equation}
    \begin{split}
      \psi^{\text{o}}_{i,j} = 0 \quad (i\in N, j=1,3,5,7,9,10).
    \end{split}
  \end{equation}
  \Cref{eq:oddwf} can be reorganized to show the relations of wave functions between two neighboring unit cells. 
  With Equations~(\ref{eq:oddwf1})~$\sim$~(\ref{eq:oddwf3}), we have 
  \begin{align*}
    \psi^{\text{o}}_{i,6} &= -\psi^{\text{o}}_{i-1,8}-\psi^{\text{o}}_{i,4} \\
                            &= \psi^{\text{o}}_{i-1,6}-\psi^{\text{o}}_{i,4} \\
                            &= -\psi^{\text{o}}_{i,2}-2\psi^{\text{o}}_{i,4}. \numberthis \label{eq:oddwf3v2}
  \end{align*}
  With \Cref{eq:oddwf1,eq:oddwf2,eq:oddwf4}, we have 
  \begin{align*}
    \psi^{\text{o}}_{i+1,2} &= -\psi^{\text{o}}_{i-1,8} \\
                            &= \psi^{\text{o}}_{i-1,6} \\
                            &= -\psi^{\text{o}}_{i,2}-\psi^{\text{o}}_{i,4}. \numberthis \label{eq:oddwf4v2}
  \end{align*}
  With \Cref{eq:oddwf1,eq:oddwf3v2,eq:oddwf4v2}, we have 
  \begin{align*}
    \psi^{\text{o}}_{i+1,4} &= -\psi^{\text{o}}_{i-1,6} - \psi^{\text{o}}_{i+1,2} \\
                            &= (\psi^{\text{o}}_{i,2}+2\psi^{\text{o}}_{i,4}) + (\psi^{\text{o}}_{i,2}+\psi^{\text{o}}_{i,4}) \\
                            &= 2\psi^{\text{o}}_{i,2}+3\psi^{\text{o}}_{i,4}. \numberthis \label{eq:oddwf1v2}
  \end{align*}
  With Euqations~(\ref{eq:oddwf3v2})~$\sim$~(\ref{eq:oddwf1v2}), the nonzero wave functions in \Cref{eq:oddwf} can be rewritten as
  \begin{subequations}\label{eq:oddwfv2}
    \begin{align}
      \begin{pmatrix}
        \psi^{\text{o}}_{i+1,2} \\ \psi^{\text{o}}_{i+1,4}
      \end{pmatrix} &= 
      \begin{pmatrix}
        -1 & -1 \\ 2 & 3 
      \end{pmatrix} 
      \begin{pmatrix}
        \psi^{\text{o}}_{i,2} \\ \psi^{\text{o}}_{i,4}
      \end{pmatrix}, \label{eq:oddwfv21mat} \\ 
      \psi^{\text{o}}_{i,6} &= -\psi^{\text{o}}_{i,2}-2\psi^{\text{o}}_{i,4}, \label{eq:oddwfv22} \\ 
      \psi^{\text{o}}_{i,8} &= \psi^{\text{o}}_{i,2}+2\psi^{\text{o}}_{i,4}, \label{eq:oddwfv23}
    \end{align} 
  \end{subequations}
  for $i\in N_+$, where \Cref{eq:oddwfv21mat} shows the relation of wave functions between two neighboring unit cells. 
  The eigenvalue of the matrix in \Cref{eq:oddwfv21mat} is the ratio of wave functions between two neighboring unit cells:
  \begin{align*}
    \lambda_1 = 1-\sqrt{2}, \quad \lambda_2 = 1+\sqrt{2},
  \end{align*}
  where $\lambda_2>1$ is unphysical for edge states.
  The eigenvector associated with $\lambda_1$ is $\left(\sqrt{2}+1,-\sqrt{2}\right)^\mathsf{T}$, and combining it with \Cref{eq:oddwfv22,eq:oddwfv23} we can fix the wave functions within the same unit cell $i$:
  \begin{equation}
    \psi^{\text{o}}_{i,2} : \psi^{\text{o}}_{i,4} : \psi^{\text{o}}_{i,6} : \psi^{\text{o}}_{i,8} =
     (\sqrt{2}+1):(-\sqrt{2}):(\sqrt{2}-1):(1-\sqrt{2}), \quad i\in N_+.
  \end{equation}
  This relation also indicates $\psi^{\text{o}}_{1,2}:\psi^{\text{o}}_{1,4}=x:(-1-x)=(\sqrt{2}+1):(-\sqrt{2})$, which gives
  \begin{equation}
    x=-\sqrt{2}-1.
  \end{equation}
  Substituting the value of $x$ into the upper panel of \Cref{fig:EdgeState}D and considering the decay ratio $\lambda_1$, the analytical solution for the edge state can be obtained, as shown in the upper panel of \Cref{fig:EdgeState}B with a normalization of wave function.
  We can see that the zero-energy edge mode with odd mirror symmetry dominates on the concave region, which explains the distribution of wave function of the HHGY in the upper panel \Cref{fig:EdgeState}A.

  In order to obtain the analytical solution of the zero-energy edge mode with even mirror symmetry, we draw the schematic structure for derivation in the lower panel of \Cref{fig:EdgeState}C.
  Similarly, we only need to consider the wave functions of ten sites in one unit cell due to the mirror symmetry.
  For the zero energy state wtih even mirror symmetry, we have
  \begin{equation}
    \psi^{\text{e}}_{i,2} = \psi^{\text{e}}_{i,9} = 0, \quad i \in N
  \end{equation}
  with the superscript ``e'' denoting even mirror symmetry and $i$ labeling the unit cell.
  Starting with $\psi^{\text{e}}_{0,1}=1$, the relations of wave functions at the following sites can be derived for the zero-energy mode:
  \begin{equation}
    \begin{split}
      & (\text{site}~0,2)\quad \psi^{\text{e}}_{0,3} = -\psi^{\text{e}}_{0,1} = -1, \\
      & (\text{site}~0,3)\quad \psi^{\text{e}}_{0,4} = -\psi^{\text{e}}_{0,2} = 0, \\
      & (\text{site}~0,4)\quad \psi^{\text{e}}_{0,5} = -\psi^{\text{e}}_{0,3} = 1, \\
      & (\text{site}~0,5)\quad \psi^{\text{e}}_{0,6} = -\psi^{\text{e}}_{0,4} = 0. \\
    \end{split}
  \end{equation}
  Similarly, starting with $\psi^{\text{e}}_{1,1}=y$, we have the relations of wave functions at the following sites:
  \begin{equation}\label{eq:evenwfuc1}
    \begin{split}
      & (\text{site}~1,2)\quad \psi^{\text{e}}_{1,3} = -\psi^{\text{e}}_{1,1} = -y, \\
      & (\text{site}~0,6)\quad \psi^{\text{e}}_{0,7} = -\psi^{\text{e}}_{0,5}-\psi^{\text{e}}_{1,3} = y-1, \\
      & (\text{site}~1,3)\quad \psi^{\text{e}}_{1,4} = -\psi^{\text{e}}_{0,6}-\psi^{\text{e}}_{1,2} = 0, \\
      & (\text{site}~0,7)\quad \psi^{\text{e}}_{0,8} = -\psi^{\text{e}}_{0,6} = 0, \\
      & (\text{site}~1,4)\quad \psi^{\text{e}}_{1,5} = -\psi^{\text{e}}_{1,3} = y, \\
      & (\text{site}~0,8)\quad \psi^{\text{e}}_{0,9} = -\psi^{\text{e}}_{0,7}-\psi^{\text{e}}_{1,5} = 1-2y. \\
    \end{split}
  \end{equation}
  Since we already know $\psi^{\text{e}}_{0,9}=0$, combining it with the last line of \Cref{eq:evenwfuc1} we obtain
  \begin{equation}\label{eq:y}
    y=1/2
  \end{equation}
  For site 5 in 1st unit cell and site 9 in 0th unit cell, the relations are
  \begin{equation}
    \begin{split}
      & (\text{site}~1,5)\quad \psi^{\text{e}}_{1,6} = -\psi^{\text{e}}_{0,8}-\psi^{\text{e}}_{1,4} = 0, \\
      & (\text{site}~0,9)\quad \psi^{\text{e}}_{0,10} = -\psi^{\text{e}}_{0,8}-\psi^{\text{e}}_{2,2} = 0. \\
    \end{split}
  \end{equation}
  Finally, setting $\psi^{\text{e}}_{2,1}=z$, we have the relations of wave functions at the following sites:
  \begin{equation}\label{eq:evenwfuc2}
    \begin{split}
      & (\text{site}~2,2)\quad \psi^{\text{e}}_{2,3} = -\psi^{\text{e}}_{0,9}-\psi^{\text{e}}_{2,1} = -z, \\
      & (\text{site}~1,6)\quad \psi^{\text{e}}_{1,7} = -\psi^{\text{e}}_{1,5}-\psi^{\text{e}}_{2,3} = z-y, \\
      & (\text{site}~2,3)\quad \psi^{\text{e}}_{2,4} = -\psi^{\text{e}}_{1,6}-\psi^{\text{e}}_{2,2} = 0, \\
      & (\text{site}~1,7)\quad \psi^{\text{e}}_{1,8} = -\psi^{\text{e}}_{1,6} = 0, \\
      & (\text{site}~2,4)\quad \psi^{\text{e}}_{2,5} = -\psi^{\text{e}}_{2,3} = z, \\
      & (\text{site}~1,8)\quad \psi^{\text{e}}_{1,9} = -\psi^{\text{e}}_{1,7}-\psi^{\text{e}}_{2,5} = y-2z. \\
    \end{split}
  \end{equation}
  Combining the known $\psi^{\text{e}}_{1,9}=0$ and the last line of \Cref{eq:evenwfuc2}, we obtain
  \begin{equation}\label{eq:z}
    z=y/2=1/4.
  \end{equation} 
  From \Cref{eq:evenwfuc1,eq:evenwfuc2,eq:y,eq:z} we can find the decay ratio of wave funtions between two neighboring unit cells are $z/y=1/2$. 
  With all the paramters obtained above, the analytical solution for the zero-energy mode with even mirror symmetry can be derived, as shown in the lower panel of \Cref{fig:EdgeState}B, where a normalization of wave function is considered.
  The zero-energy edge state with even mirror symmetry dominates on the outmost carbon chain of the edge, which explains the distribution of wave function in the lower panel of \Cref{fig:EdgeState}A.

 %\bibliography{TopoHGY}

\providecommand{\latin}[1]{#1}
\makeatletter
\providecommand{\doi}
  {\begingroup\let\do\@makeother\dospecials
  \catcode`\{=1 \catcode`\}=2 \doi@aux}
\providecommand{\doi@aux}[1]{\endgroup\texttt{#1}}
\makeatother
\providecommand*\mcitethebibliography{\thebibliography}
\csname @ifundefined\endcsname{endmcitethebibliography}
  {\let\endmcitethebibliography\endthebibliography}{}

\end{document}